\documentclass[twocolumn, aps, showpacs] {revtex4}
\usepackage{graphicx}
\begin{document}
\draft

\title {Spin polarization and g-factor of a dilute GaAs two-dimensional electron
system}

\author {E. Tutuc, S. Melinte and M. Shayegan}
\address{Department of Electrical Engineering, Princeton
University, Princeton, NJ 08544}
\date{\today}
\begin{abstract}
The effective g-factor ($g^{*}$) of a dilute interacting
two-dimensional electron system is expected to increase with
respect to its bare value as the density is lowered, and to
eventually diverge as the system makes a transition to a
ferromagnetic state. We report here measurements of $g^{*}$ in
dilute (density 0.8 to 6.5 $\times10^{10}$cm$^{-2}$),
high-mobility GaAs two-dimensional electrons from their spin
polarization in a parallel magnetic field. The data reveal a
surprising trend. While $g^{*}$ is indeed significantly enhanced
with respect to the band g-factor of GaAs, the enhancement factor
{\it decreases} from about 6 to 3 as the density is reduced.
\end{abstract}
\maketitle

The ground state of a dilute, interacting electron system has been
of interest for decades. It has long been expected that, because
of interaction, such a system makes a transition to a
ferromagnetic state as the density is reduced below a certain
threshold \cite{bloch, stoner}. For even lower densities, the
system should eventually become an electron solid (Wigner crystal)
\cite{wigner}. A relevant parameter associated with this evolution
is the effective g-factor ($g^{*}$) of the system. In the limit of
high density, when the parameter $r_{s}$, the average
interparticle separation   measured in units of effective Bohr
radius, approaches zero, $g^{*}$ should have the "bare" value
determined by the energy band structure of the host material. With
decreasing density, $g^{*}$ is expected to increase monotonically
and diverge at the density below which the electron system enters
its ferromagnetic state. For an ideal two-dimensional electron
system (2DES), quantum Monte Carlo calculations \cite{tanatar,
varsano} indeed confirm the above trend.

An excellent candidate for testing these predictions is the GaAs
2DES, as it possesses very low disorder combined with a simple
band structure. Here we report measurements of the spin
polarization of a very high quality, dilute 2DES in a modulation
doped GaAs/AlGaAs heterostructure as a function of an in-plane
magnetic field. Via transport measurements, we find the magnetic
field above which the 2DES becomes fully spin polarized, and from
this field we determine $g^{*}$. The results reveal a remarkable
trend: as the density is lowered from 6.5 to 0.8
$\times10^{10}$cm$^{-2}$, corresponding to an increase in $r_{s}$
from 2.1 to 6.3, the measured $g^{*}$ decreases from 2.7 to 1.3.
This implies a substantial overall enhancement of $g^{*}$ with
respect to the band g-factor of GaAs ($|g_{b}|=0.44$ in GaAs
\cite{weisbuch}). The decrease of $g^{*}$ with $r_{s}$, however,
is unexpected.

We studied a Si-modulation doped GaAs heterostructure grown on a
(100) GaAs substrate. We used a square sample in a Van der Pauw
geometry, with a backgate to control the density. We made
measurements in a dilution refrigerator at a temperature ($T$) of
$\simeq25$mK and magnetic fields ($B$) up to 18T, and in pumped
$^{3}He$ at $\simeq0.3$K and fields up to 33T. The sample was
mounted on a single-axis tilting stage that can be rotated, using
a computer controlled stepper motor, in order to change the angle
($\theta$) between the sample plane and the magnetic field. The
measurements were done using low-frequency lock-in techniques. At
zero gate bias, the sample has a density $n=1.4\times10^{10}$
cm$^{-2}$ and a mobility of 55 m$^{2}$/Vs.

An external magnetic field applied parallel to the 2DES causes a
Zeeman splitting of the energy bands. This splitting induces a
difference in population of the spin-up and spin-down subbands,
which leads to a net spin polarization of the system. If the
splitting exceeds the Fermi energy of the system, all spins are
aligned and the 2DES is fully spin polarized. Assuming a simple
model, in which $g^{*}$ is independent of the applied magnetic
field, we can write the splitting between the spin-up and
spin-down subbands as $E_{Z}=|g^{*}|\mu_{B}B$, where $\mu_{B}$ is
the Bohr magneton. In this model the 2DES becomes fully spin
polarized at a field $B_{P}$, given by
\begin{equation}\label{relation}
  B_{P}=(h^{2}/2\pi\mu_{B})\cdot(n/m^{*}g^{*})
\end{equation}
where $m^{*}$ is the effective mass, $n$ is the total density of
the 2DES and $h$ is Planck's constant. We emphasize that  we
measure the effective g-factor defined, as in Eq. (1), by the
field at which full polarization is achieved.

In our experiments we measure, via the analysis of Shubnikov-de
Haas (SdH) oscillations, the Fermi contours of the two spin
subbands. We apply a constant magnetic field parallel to the 2D
plane and slowly rotate the sample around $\theta=0^{\circ}$ to
induce a small perpendicular field component, $B_{\bot}$. We
record the sample resistance during the rotation, and Fourier
analyze its SdH oscillations with $B_{\bot}$ to obtain the
populations of the two spin subbands \cite{tutuc}. Our experiments
allow a determination of the field, $B_{P}$, above which the
minority spin-subband depopulates and the 2DES becomes fully spin
polarized. In the range where the Fourier transforms are done the
parallel component of the field, $B_{\|}$, is equal to the total
field $B$ to better than 2\%.

In Fig. 1 we show plots of the sample resistance $R$ vs.
$B_{\bot}$, taken at a density $n=2.05\times10^{10}$ cm$^{-2}$, as
determined from the positions of quantum Hall states. The top
trace was taken in a purely perpendicular field. The Fourier
transform (FT), shown on the right, exhibits two peaks, one at
0.85T and another at approximately half this value, 0.42T. The
0.85T frequency, when multiplied by $(e/h)$, gives
$2.05\times10^{10}$ cm$^{-2}$, i.e., the total density of the
2DES. The 0.42T peak stems from the spin unresolved SdH
oscillations. The rest of the traces shown in Fig. 1 were taken by
rotating the sample at the indicated $B$ applied almost parallel
to the 2DES. With increasing $B_{\|}$, we observe a splitting of
the lower FT peak (0.42T) into two peaks. The positions of these
two peaks, multiplied by $(e/h)$, give the two spin subband
populations. Note that the two populations add up to the total
density of the sample. As $B_{\|}$ is increased, the majority spin
subband peak merges with the total density peak (0.85T) and the
minority spin subband peak moves to very low frequencies and is no
longer resolved \cite{R2}.

\begin{figure}
\centering
\includegraphics[scale=0.42]{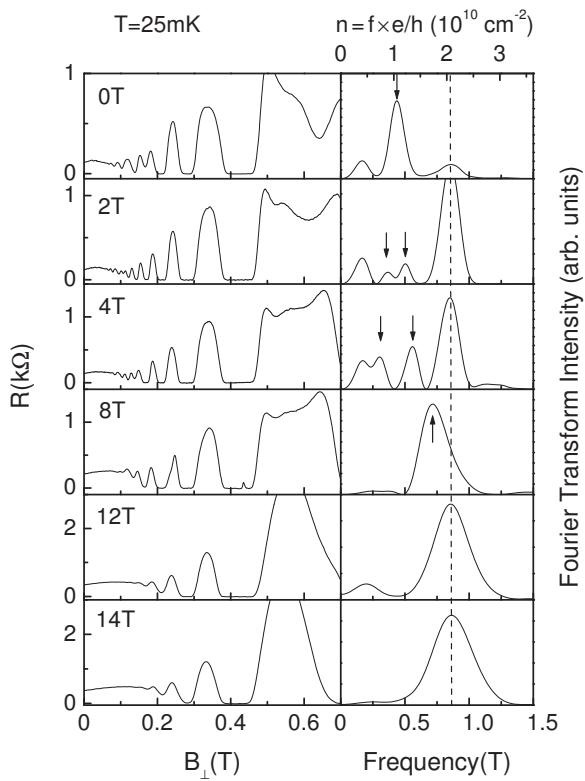}
\caption {\small{Resistance vs. perpendicular field at the
indicated parallel fields and the Fourier transforms of the SdH
oscillations. The total density of the 2DES is $2.05\times10^{10}$
cm$^{-2}$.}}
\end{figure}

In Fig. 2 we summarize the positions of the FT peaks corresponding
to the majority and minority spin subbands as a function of
$B$ for the case examined in Fig. 1. Above a certain field
$B_{P}$ the majority spin subband population saturates at a value
which corresponds to the total density of the 2DES. Therefore,
$B_{P}$ marks the onset of full spin polarization. Within the
experimental error, the evolution of the FT peak positions as a
function of field is linear. This implies that $g^{*}$ is roughly
independent of the applied parallel field.

We also measured the in-plane magnetoresistance (MR), by fixing
$\theta$ at $0^{\circ}$ and recording the resistance as a function
of the applied magnetic field. The MR trace, taken for
$n=2.05\times10^{10}$ cm$^{-2}$, is also shown in Fig. 2. This
trace exhibits a clear break in the functional form of the MR as
it changes from an $\sim e^{B^{2}}$ dependence at low field to a
simple exponential, $\sim e^{B}$ dependence at higher fields. The
data in Fig. 2 demonstrate that the onset of the simple
exponential behavior of the in-plane MR coincides with the field
$B_{P}$ above which the spins are fully polarized. Remarkably, the
same functional behavior of the in-plane MR is seen in GaAs 2D
hole systems \cite{yoon}, and it has been shown \cite{tutuc} that
the onset of the simple exponential regime corresponds to the full
spin polarization of the 2D system. Several studies on 2D
electrons in Si-MOSFETs have also pointed out a correlation
between the in-plane MR and the full spin polarization
\cite{mertes,okamoto,vitkalov,dolgopolov}.

\begin{figure}
\centering
\includegraphics[scale=0.36]{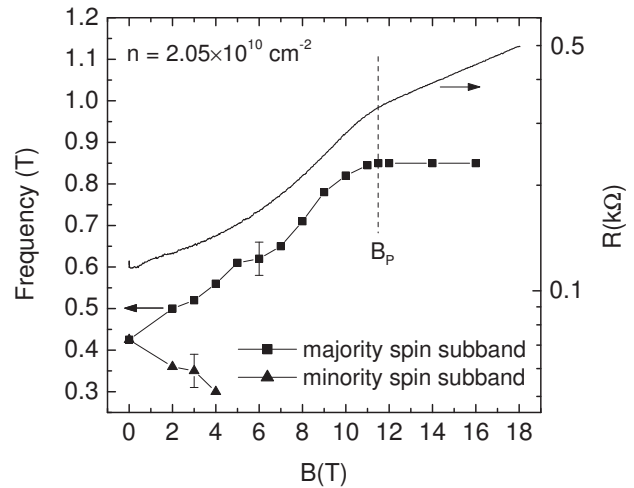}
\caption {\small{Summary of FT peak positions as a function of
applied magnetic field, along with the in-plane magnetoresistance
trace. The dashed line marks the field $B_{P}$ at which the 2DES
becomes fully spin polarized.}}
\end{figure}

The strong correlation between the full spin polarization and the
onset of exponential MR provides another method of finding the
field $B_{P}$. While this method does not allow a direct
measurement of the spin subband populations, it can be useful at
lower densities where the SdH method is no longer practical
because of a decrease in the number of resistance oscillations. We
summarize in Fig. 3 our MR data taken at different densities. At
all densities, the MR exhibits an $\sim e^{B}$ dependence at high
fields and the onset of this dependence clearly depends on the
density. We emphasize that in several cases where we have made
both MR and SdH oscillations measurements in constant field, the
values of $B_{P}$ obtained from the two methods coincide.

\begin{figure}
\centering
\includegraphics[scale=0.37]{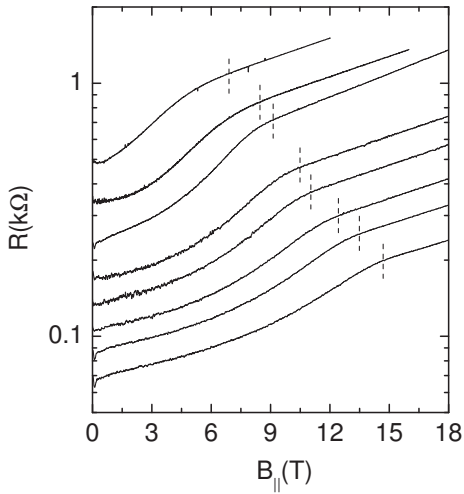}
\caption {\small{In-plane magnetoresistance at different
densities, from top to bottom: 0.8, 1.08, 1.4, 1.68, 1.9, 2.18,
2.47, 2.89$\times10^{10}$ cm$^{-2}$. Dashed lines mark the field
above which the magnetoresistance exhibits a simple exponential
dependence. Top two traces were taken at a temperature of 0.3K,
the rest at 25mK. Top trace was divided by a factor of 3.}}
\end{figure}

Using the measured values of $B_{P}$ and the relation (1) we
determine $g^{*}$. For $B_{P}$ we use the saturation field of the
majority spin subband FT peak and the onset of the exponential
regime of the in-plane MR. Another parameter needed is the
effective mass, $m^{*}$, which also can be different from the band
effective mass due to electron-electron interaction \cite{kwon}.
We independently measured $m^{*}$ from the $T$ dependence of SdH
oscillations in a purely perpendicular magnetic field at several
densities in the range 1.4 to 3$\times 10^{10}$ cm$^{-2}$. The
analysis uses the Dingle formula, $\Delta R/R_{0} \sim
\xi/\sinh\xi$, where $\Delta R/R_{0}$ is the normalized amplitude
of the SdH oscillations, $\xi\equiv2\pi^{2}k_{B}T/\hbar\omega_{c}$
and $\omega_{c}=eB/m^{*}$. Fits of the Dingle formula to the data
are shown in Fig. 4, together with the values of $m^{*}$ that
provide the best least-squares fit. Within the experimental error
(7\%), $m^{*}$ is the same as the band effective mass of electrons
in GaAs, $m_{b}=0.067m_{e}$, where $m_{e}$ is the free electron
mass. For simplicity, in determining $g^{*}$ we have used the band
effective mass.

\begin{figure}
\centering
\includegraphics[scale=0.35]{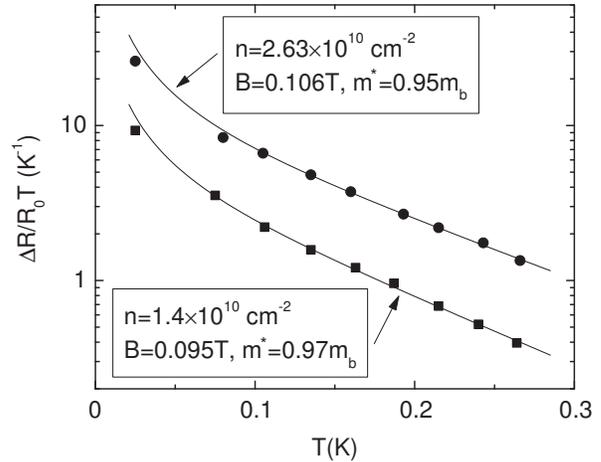}
\caption{\small{Dingle plots of $\Delta R/R_{0}T$ vs. $T$. The
density and the magnetic field at which the resistance was
measured are indicated. The lines represent least-squares fits to
the Dingle formula. For each dataset, $m^{*}$ obtained from the
fit is shown.}}
\end{figure}

In Fig. 5 we plot the values of $g^{*}$, normalized to the GaAs
band g-factor, as a function of density and $r_{s}$ \cite{jiang}.
The open and closed symbols denote the two methods used to
determine $B_{P}$; their overlap confirms that the onset of
exponential regime of the MR marks the full spin polarization. For
the lowest density traces, we have data available only at 0.3K.
However, $T$ dependence of the in-plane MR, measured at
$n=1.4\times10^{10}$ cm$^{-2}$, showed only a small variation of
$B_{P}$ with $T$ in the range 25mK to 0.5K. The larger error bars
for the lowest density points in Fig. 5 take into account such
variation with $T$.

Figure 5 data highlight the main finding of our work: while the
measured values of $g^{*}$ are three to six times larger than the
GaAs band g-factor, this enhancement {\it decreases} with
increasing $r_{s}$. This trend is unexpected and at variance with
theoretical calculations which predict that, for an ideal 2DES,
$g^{*}$ should increase as $r_{s}$ increases and the interaction
becomes more important \cite{tanatar,varsano}. While we do not
have an explanation for this discrepancy, we point out that at
least three factors distinguish our samples from ideal 2D systems:
finite layer thickness, disorder, and the spin-orbit interaction.

For a 2DES with finite layer thickness, even in a single-electron
picture, an in-plane magnetic field can affect the energy bands
and therefore both the effective mass \cite{wieck} and the
g-factor. Indeed, band calculations \cite{winkler} performed for
the 2DES studied in our work reveal some enhancement of the
effective mass and, to a lesser degree, a reduction of the
g-factor with in-plane field. However, these modifications of the
band parameters appear to be too small to explain the data of Fig.
5.

The effect of disorder is more subtle. As the density is reduced,
the disorder potential can play a more dominant role than the
electron-electron interaction, and may lead to an inhomogeneous
spatial distribution of the electrons in the sample. If the
electrons become localized in the potential minima, it is possible
that the behavior of the 2DES reverts to that of a single-particle
system and $g^{*}$ decreases. Such a scenario is consistent with
the data of Fig. 5: for $n=0$, $g^{*}/g_{b}$ extrapolates to a
value close to unity. On the other hand, we have made $g^{*}$
measurements on a sample with a mobility three times lower than
the present specimen, and obtained similar results as those shown
in Fig. 5. Moreover, our sample exhibits no signs of strong
electron localization: it exhibits SdH oscillations, and its
resistivity is less than $h/e^{2}$.

\begin{figure}
\centering
\includegraphics[scale=0.36]{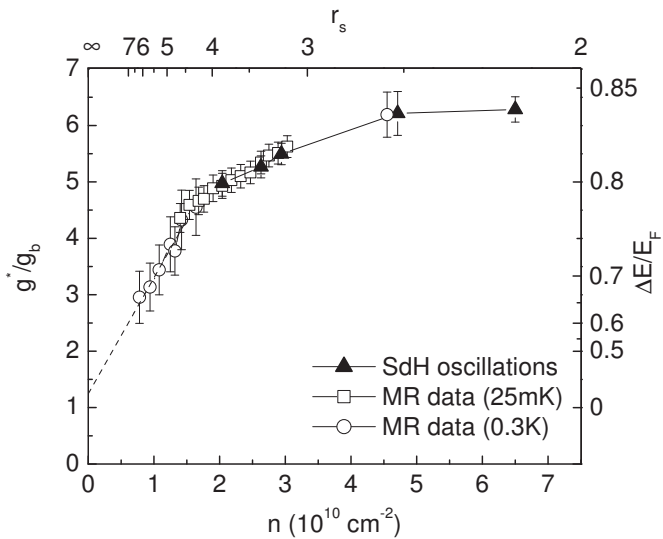}
\caption {\small{The effective g-factor normalized to the GaAs
band g-factor ($|g_{b}|=0.44$) vs. $n$ or $r_{s}$. The symbols
denote the two different ways of measuring the full spin
polarization field, $B_{P}$: from SdH oscillations in constant
field (closed symbols) or from the onset of the exponential
behavior of the magnetoresistance (open symbols). The dashed line
is a linear fit to the lower density data points. On the right
scale we indicate the corresponding values of $\Delta E/E_{F}$, as
defined in the text.}}
\end{figure}

An examination of the results of several studies
\cite{okamoto,pudalov,vitkalov} which have reported measurements
of $g^{*}$ in Si-MOSFET 2D electrons provides further argument
against disorder being responsible for the anomalous behavior we
observe in Fig. 5. These studies have generally reported an
enhancement of $g^{*}$ with increasing $r_{s}$ in much the same
range of $r_{s}$ that we have examined. The mobility of our 2DES,
however, is about a factor of 10 larger than in Si-MOSFETs even
though our densities are about 10 times smaller, implying lower
disorder in our samples. If disorder were the main culprit, one
would expect that Si-MOSFET data should also show a decreasing
$g^{*}$ with increasing $r_{s}$ \cite{R3}.

Another important observation is worth describing. As we mentioned
before, the in-plane MR of GaAs 2D {\it holes} was recently
reported \cite{yoon,papadakis,tutuc}, and a relation between the
MR behavior and full spin polarization, similar to that shown in
Fig. 2, was established \cite{tutuc}. If we convert the magnetic
fields above which the 2D holes are fully spin polarized to an
effective $g^{*}$, we find a trend very similar to the one seen in
Fig. 5: $g^{*}$ is enhanced with respect to its band value but it
decreases with increasing $r_{s}$. We emphasize that the GaAs 2D
holes too have very high mobilities and should contain low
disorder. We conclude that, while we cannot rule out the role of
disorder, the unexpected behavior we observe in Fig. 5 appears to
be intrinsic to low disorder GaAs 2D electrons and holes.

The spin-orbit interaction, present in both GaAs 2D electron and
hole systems, but nearly absent in Si-MOSFETs, may also play a
role here. Since $g_{b}$ is significantly influenced by the
spin-orbit interaction, one may expect that this interaction
should modify $g^{*}$ in a many-body picture also. It is not
clear, however, how the spin-orbit interaction would explain the
trend in Fig. 5.

Finally, in Fig. 5 we provide a measure of the interaction energy
for the spin polarization of our 2DES. We introduce the
enhancement energy $\Delta E$, defined as the difference between
the band Fermi and Zeeman energies, both evaluated at the in-plane
field where the 2DES becomes fully spin polarized, i.e. $\Delta
E=(2\pi \hbar^{2}/m_{b})n-g_{b}\mu_{B}B_{P}$. The energy $\Delta
E$, measured in units of the Fermi energy is then simply equal to
$(1-g_{b}/g^{*})$. This quantity, which is indicated on the right
scale of Fig. 5, reiterates the main result of our study. The
measured $\Delta E/E_{F}$ decreases with increasing $r_{s}$, while
theoretical calculations \cite{tanatar,varsano} predict the
opposite: $\Delta E/E_{F}$ should be zero at $r_{s}=0$, increase
monotonically with $r_{s}$, and reach unity for $r_{s}$ larger
than a critical value of $\simeq30$.

We thank E.P. De Poortere, D.M. Ceperley, A.H. MacDonald, B.L.
Altshuler and R. Winkler for fruitful discussions. This work was
supported by the DOE, NSF and the von Humboldt foundation. Part of
the work was done at NHMFL which is supported by NSF; we also
thank T. Murphy and E. Palm.


\begin{thebibliography}{10}
\bibitem{R2}
The peaks observed at very low frequencies ($<0.2$T) in Fig. 1 are
FT artifacts.
\bibitem{R3}
There is, however, a difference between the disorder potentials in
modulation doped heterostructures and Si-MOSFETs: the former is
dominated by long range fluctuations while the latter by short
range fluctuations.
\end{thebibliography}
\end{document}